\documentclass[aps,amsmath,twocolumn,amssymb,showkeys,showpacs,10pt]{revtex4-1}

\usepackage{graphicx}
\usepackage{amssymb}
\usepackage{amsmath}
\usepackage{bm}
\usepackage{enumitem}

\usepackage{graphicx}
\usepackage{amssymb}
\usepackage{amsmath}
\usepackage{bm}
\usepackage{enumitem}

\usepackage{graphicx}
\usepackage{dcolumn}
\usepackage{bm}
\usepackage{hyperref}
\usepackage{amssymb}
\usepackage{hyperref}
\usepackage{graphicx}
\usepackage{hyperref}
\hypersetup{colorlinks,breaklinks,
            urlcolor=[rgb]{0.,0.4,0.4},
            linkcolor=blue,
            citecolor=red}
\usepackage{charter}
\providecommand{\U}[1]{\protect\rule{.1in}{.1in}}
\begin{document}

\title{Decay properties  of unstable Tonks-Girardeau gases from  a     split trap    }
\author{Przemys\l aw Ko\'{s}cik}
\affiliation{ Department of Computer Sciences, State Higher Vocational School in
Tarn\'{o}w, ul. Mickiewicza 8, PL-33100 Tarn\'{o}w, Poland}
\email{p$_$koscik@pwsztar.edu.pl}
\begin{abstract}

We study the decay properties of Tonks-Girardeau gases escaping from a double-well potential. Initially, the gases are constrained between two infinite $\delta$ barriers with an on-center $\delta$-split potential. The strength of one of the obstacles is suddenly reduced, and the particles start to tunnel to the open space.
Using the resonance expansion method, we derive the single-term   approximate expression for the $N$-particle survival probability  and   demonstrate its effectiveness in both the exponential and nonexponential regimes. We also predict a parity effect and provide physical insights into its nature at different stages of the time evolution. 
 We conclude that only the initial phase  of the decay of the many-particle state may comply with an exponential law. The decay properties are dramatically affected by the presence of the split barrier. Our results reveal  the overall decay mechanism of unstable Tonks-Girardeau gases from single and double quantum wells.

\end{abstract}

\maketitle

\newcommand{\appropto}{\mathrel{\vcenter{
  \offinterlineskip\halign{\hfil$##$\cr
    \propto\cr\noalign{\kern2pt}\sim\cr\noalign{\kern-2pt}}}}}
\section{Introduction}

 Since the birth of quantum physics, there has been much interest in
the properties of particles tunneling from an open well trap \cite{alpha,alpha1,alpha2,alpha3}. Many research studies have been conducted so far in this context,  and various models have been considered in the literature \cite{to,Koide,pons,Romo,urb,garcia}. In particular, the
 so-called \textit{Winter model} \cite{Winter0}, consisting  of an infinite wall and a  $\delta$-\textit{function}  potential,  has attracted significant attention \cite{Winter1,Winter2,Winter3,TG0,TG1,kkl}. A good feature of the \textit{Winter's model}  is that its energy eigenfunctions can be found in closed analytical forms, enabling us to gain insights into the properties of the decaying particles easily.
 With this concrete example,   Winter \cite{Winter0} showed   that the decay process
 exhibits deviations from the exponential law, and its long-time evolution follows a power law.
 The nonexponential decay has been observed in a variety of experiments
\cite{wil,wil1,exp}.
New advances in experimental techniques have provided  opportunities to study tunneling phenomena in a controllable way \cite{exp,johim1,johim2}, thereby accelerating research activity towards a better understanding of the properties of unstable many-body states.
 For instance,    the recent studies include systems of two identical noninteracting particles \cite{two}, ultracold-atom systems  \cite{sow,sow1,Lode,Lode1,Blume,Brand,in,in1},
two-particle systems with Coulomb interactions \cite{Maru1}, and a model system consisting of a core nucleus and two valence protons \cite{Maru}. It is worth mentioning that the complex scaling method \cite{Moj} also provides a tool for studying the resonance states of many-particle systems \cite{ku0,ku1,ku2,ku3}. However,  few attempts have been made to increase our understanding of the decay properties of systems with more than two particles.

One of the essential results regarding the one-dimensional (1D) systems is the famous Bose-Fermi (BF) mapping \cite{Girardeau}. In the most straightforward case, the BF  mapping   relates the wave function of bosons with infinitely strong $\delta$ repulsions (Tonks-Girardeau gas) to a free-fermion gas.
 The BF mapping also holds  for the time-dependent case \cite{Girardeau1}. As a result, the theoretical investigations of unstable  Tonks-Girardeau (TG) gases can be considerably simplified, which offers unique opportunities to gain physical insights into the tunneling process at the few-body level. Some researchers have taken action  to explore the decay properties of TG gases
  \cite{pons,TG0,TG1}, where one of the  relevant results concerns the long-time  decay   \cite{TG1}.

This work aims to carry out a comprehensive investigation of
the properties of unstable TG gases escaping from an open double-well potential. Our interest in such systems is primarily   motivated by the implementation of  double-well structures in laboratories and theoretical works in this area  \cite{in1,double0,double1,double2,double3,double4,double5,in00,in01}.
A   simple candidate by which to simulate a double-well structure is  \textit{Winter's model} with an additional $\delta$-split barrier at the center of the trap.

We examine the typical scenario of the  tunneling process; namely,  the   $N$ hardcore bosons are initially confined in the  hard-wall split trap. At some time, one of the barriers is lowered, and the initial state starts to decay in time.  Within the framework of the resonance expansion approach  \cite{garcia}, we identify the tunneling mechanisms in different time regions by providing a closed-form approximate expression for the $N$-particle survival probability. Our results go beyond the long-time decay and, in this sense, they extend the results in Ref. \cite{TG1}. We also address the question of how the decay properties are influenced by the $\delta$-split barrier when changing its strength and the number of particles, $N$.

The remainder of this study is as follows.
 In Sec. \ref{section1}, we  present the model and theoretical tools used to probe its decay properties.
  Section \ref{section2}  is devoted to the features of
decaying one- and many-particle states.
    Some concluding remarks  are found in Sec. \ref{section3}.

\section{ Model}\label{section1}
The following Hamiltonian describes the 1D system of $N$  indistinguishable bosons interacting via a contact potential:

\begin{equation}{\cal \hat{H}}= \sum_{i=1}^{N}{ \hat{h}(i)}+g \sum_{j<k}\delta(x_{j}-x_{k}),\end{equation}
where  $\hat{h}$ is the single-particle Hamiltonian and  $g$ is the strength of the interaction \cite{olsh}. For the present system, the  Hamiltonian $\hat{h}$ takes the following form: \begin{equation}\label{sh}
 { \hat{h}}= -\frac{\hbar ^2}{2m}\frac{\mathrm{d}^2}{\mathrm{d}x^2}+{\cal V}(x)+\alpha\delta(x)+\eta\delta(x-L),
\end{equation} with  $\alpha\geq0$ and $\eta \geq0$, where ${\cal V}(x)$ represents a hard-wall trap that is infinite in the region $x<-L$ and  zero elsewhere, and $2L$ is the width of the trapping potential. With the emergence of new technologies, such a 1D  box split trap  can be experimentally fabricated  by laser trapping techniques.
In the  TG limit  ($g\rightarrow \infty$) that we are interested in,  the bosonic wave function  is related via  BF  mapping \cite{Girardeau} to the corresponding wave function  of spineless free fermions as follows:
\begin{equation}
\Psi(x_{1},x_{2},...,x_{N})=\hat{\Upsilon}\Psi_{F}(x_{1},x_{2},...,x_{N}),
\end{equation}
where $\hat{\Upsilon}$ is the mapping function
$\hat{\Upsilon}=\Pi_{ k<l}\mathrm{sgn}(x_{k}-x_{l})$
that ensures  the bosonic symmetry ($\mathrm{sgn}$ is the sign function).
  We choose
 the initial state ($t<0$) as  the  ground state  of $N$ hardcore bosons in a   hard-wall split trap. At   time $t=0$, the  barrier strength at  $x=L$ is instantaneously reduced to a finite value of  $\eta$,  and the initial wave function  starts to evolve with time \cite{Girardeau1}:  \label{tgfun}
\begin{eqnarray}\Psi(x_{1},x_{2},...,x_{N};t)=\nonumber\\=\Pi_{k<l}\mathrm{sgn}[x_{k}-x_{l}]{1\over \sqrt{N!}}\mathrm{det}_{i=1,j=1}^{N}[\phi_{i}(x_{j},t)], \end{eqnarray} where $\mathrm{det}$ symbolizes the determinant
and $\phi_{n}(x,0)$ is the $n$th eigenstate of the single-particle Hamiltonian (\ref{sh}) in the hard-wall limit ($\eta\rightarrow\infty$).  The evolution process of each single-particle orbital is governed by the  Schr\"{o}dinger equation:
\begin{equation}\label{sh1}
 i\hbar {\partial  \phi_{k}(x,t)\over \partial t}={\cal \hat{H}}\phi_{k}(x,t).
\end{equation}
 In the following, we take $L=\hbar= m=1$; i.e.,  the spatial and time  coordinates are measured in units of $L$ and $mL^2/\hbar$, respectively,  and energies in units of  $\hbar^2/(mL^2)$.
The  model we  consider has a convenient feature where both the continuum    wave functions  ($\eta<\infty$)  and the  eigenfunctions  of the  hard-wall  split trap  can be found in closed analytical forms (see Appendixes \ref{ap1} and \ref{ap2}). The solutions to Eq. (\ref{sh1}) maybe written  in terms of the
continuum wave functions as follows:
\begin{equation}\label{time}\phi_{k}(x,t)=\int_{0}^{\infty} C_{k}(p)\psi_p(x)e^{-{i t p^2\over 2} }\mathrm{d}p,\end{equation}
where
\begin{equation}\label{mom} C_{k}(p)=\int_{-1}^{1}{ \phi}_{k}(x,0) \psi_p(x)\mathrm{d}x.\end{equation}
The  integrals in Eq. (\ref{mom}) can be   performed explicitly  (Appendix \ref{ap3}). In contrast, those  in  Eq. (\ref{time})  need numerical integration.

 For  simplicity,  we analyze the decay properties in terms of the survival probability, also known as  quantum fidelity, as follows:
\begin{equation}S(t)=|\langle\Psi(0)|\Psi(t)\rangle|^2,\end{equation} which measures how the  time-evolved state differs from the initial state. For  the TG gases, the $N$-particle survival probability  can be simplified to the following form \cite{TG1}:
\begin{equation}\label{tg}S^{(N)}(t)=|\mathrm{det}_{k,l=1}^{N}[{\cal A}_{kl}(t)]|^2, \end{equation}
where
  ${\cal A}_{kl}(t)=\langle \phi_{k}(0)|\phi_{l}(t)\rangle$. With the use of integral representation in Eq.
  (\ref{time}),    ${\cal A}_{kl}(t)$    can be expressed by  the following 1D integrals: \begin{equation}\label{o}
{\cal A}_{kl}(t)=\int_{0}^{\infty} C_{k}^{*}(p)C_{l}(p)e^{-  {i t p^2\over 2} }\mathrm{d}p, \end{equation} which  makes   calculating  the survival probability less complicated than the nonescape probability,
  both   in the framework of the resonance expansion method \cite{garcia} and numerically.
 The integrals in Eq. (\ref{time}) are highly oscillatory; thus, their accurate  estimation  is a very tedious numerical task. One has to use the resonance expansion method to obtain insights into the decay properties   in the long-time regime, especially when $N$ is large.

 However, the survival and nonescape probabilities   have their resonance expansions in the same form \cite{garcia} (only the expansion coefficients are  different). As a result,  each characteristic  behavior
  of the former  (exponential decay, long-time decay,  etc.)  is usually  reflected    by  the latter, and vice versa \cite{pons,Romo,two}. Specifically, such behaviors of both quantities  can be described by suitable single-term approximations extracted  from their resonance expansions, which are   proportional to each other. This was adequately demonstrated in Ref. \cite{two}  for   two noninteracting particles in entangled symmetric or antisymmetric initial  states. In other words, at least in the cases of TG gases, which can be described with the resonance expansion approach, the conclusions inferred from the behaviors of the survival and nonescape probabilities are consistent in the main points.

Within the  resonance expansion approach, which is based   on  Cauchy's residue theorem,    Eq. (\ref{o}) can be rewritten in   the form

\begin{equation}\label{amp1}  {\cal A}_{kl}(t)=\sum_{r}M_{r}^{(k,l)}e^{-i t p_{r}^2/2}+M^{(k,l)}(t),\end{equation}
where $p_{r}$ are the so-called proper poles, i.e., the  roots of the denominator in Eq. (\ref{o}) on  the fourth quadrant of the complex $p$ plane, and  $M^{(k,l)}(t)$ is the   integral  contribution  along a path.   The exponential terms in Eq. (\ref{amp1}) may be rewritten  as  $e^{-i t p_{r}^2/2}=e^{-\Gamma_{r}t/2+i t\xi_{r}}$, where $\Gamma_{r}=-\mathrm{Im}[p_{r}^2]$ and  $\xi_{r}=(\mathrm{Im}[p_{r}]^2-\mathrm{Re}[p_{r}]^2)/2$   can be viewed as the decay rates  and  resonance energies, respectively.    The   positions of the proper poles  depend only   on the      system parameter values, and the same applies to $\Gamma_{r}$ and  $\xi_{r}$.
   The nonexponential component $M^{(k,l)}(t)$
   can be expanded   into a series of  inverse powers of  $t$, $\sim t^{{-n-1/2}}$ \cite{garcia,two}. Since the  integrand in Eq. (\ref{o}) is given  in  analytical form,  this   can be  achieved  explicitly  (at least term by term) \cite{two}.
    In practice,    it suffices to take into account in Eq. (\ref{amp1})  only a few terms from the poles $p_{r}$ and inverse power contributions to achieve satisfactory approximations, except for short times \cite{two1}. For details
regarding the resonance expansion approach, see Refs. \cite{garcia,TG1,two}.
From this point onwards, we assume that
$\Gamma_{r}$ are ordered according to the resonance spectra: $\xi_{1}<\xi_{2}<...<\xi_{r}<$.
 For individual states, we denote $M_{r}^{(k)}=M_{r}^{(k,k)}$ and ${\cal A}_{k}(t)={\cal A}_{kk}(t)$, and   $S_{k}(t)= |{\cal A}_{kk}(t)|^2$.

\section{Results}\label{section2}

\subsection{ Single-particle case}\label{sing}
\begin{figure}

\includegraphics[width=0.2389\textwidth]{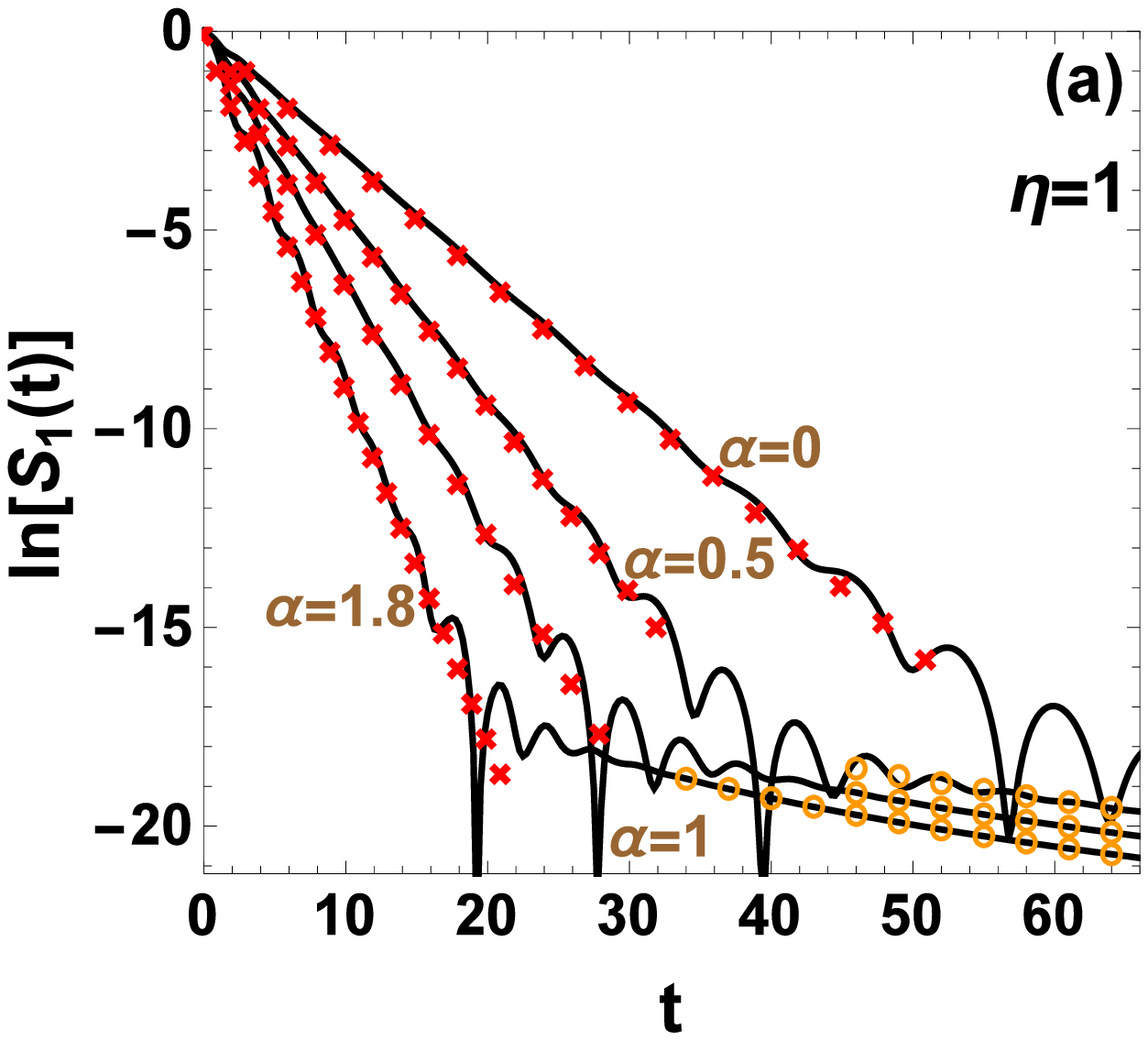}
\includegraphics[width=0.2389\textwidth]{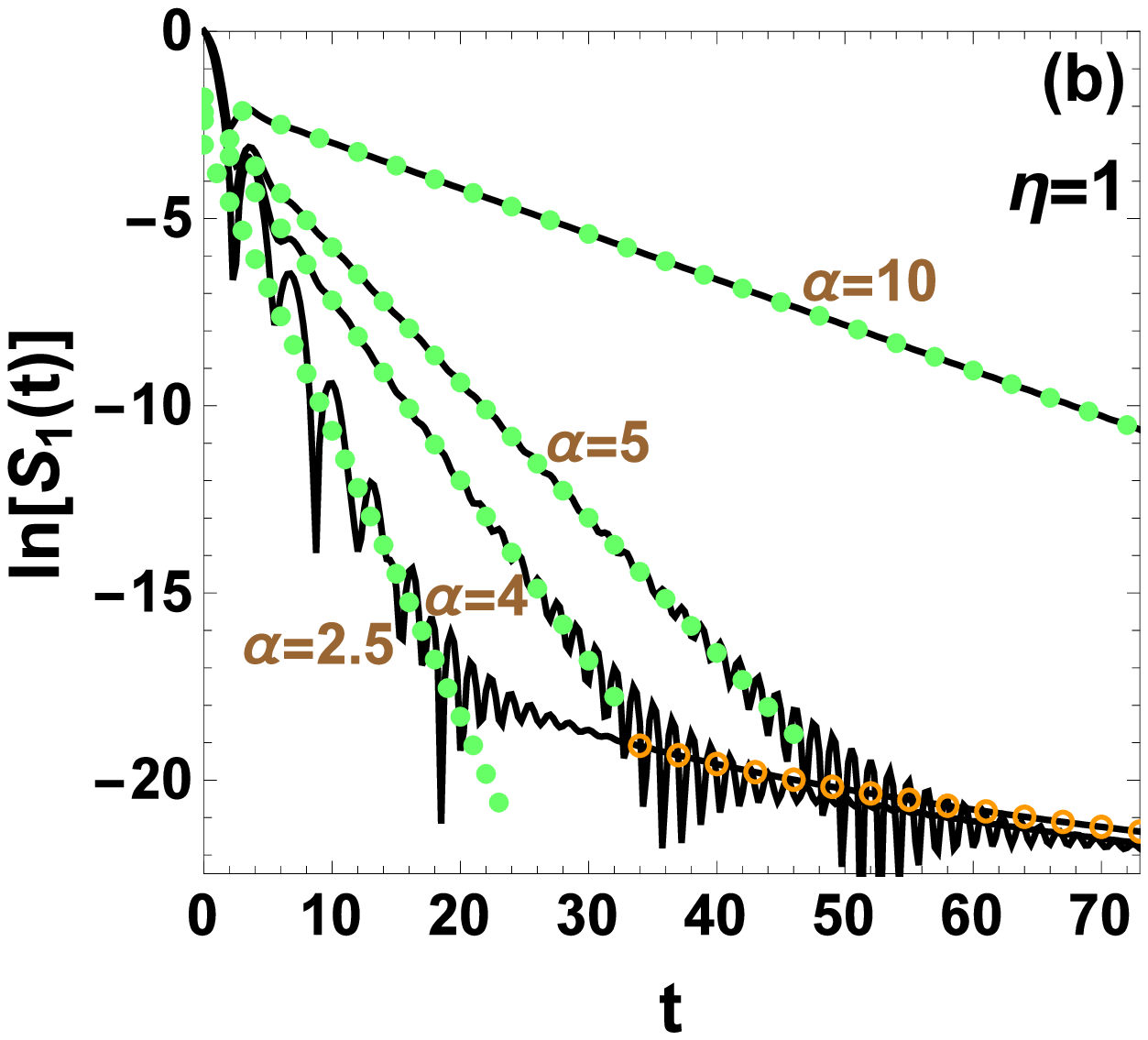}

\includegraphics[width=0.2389\textwidth]{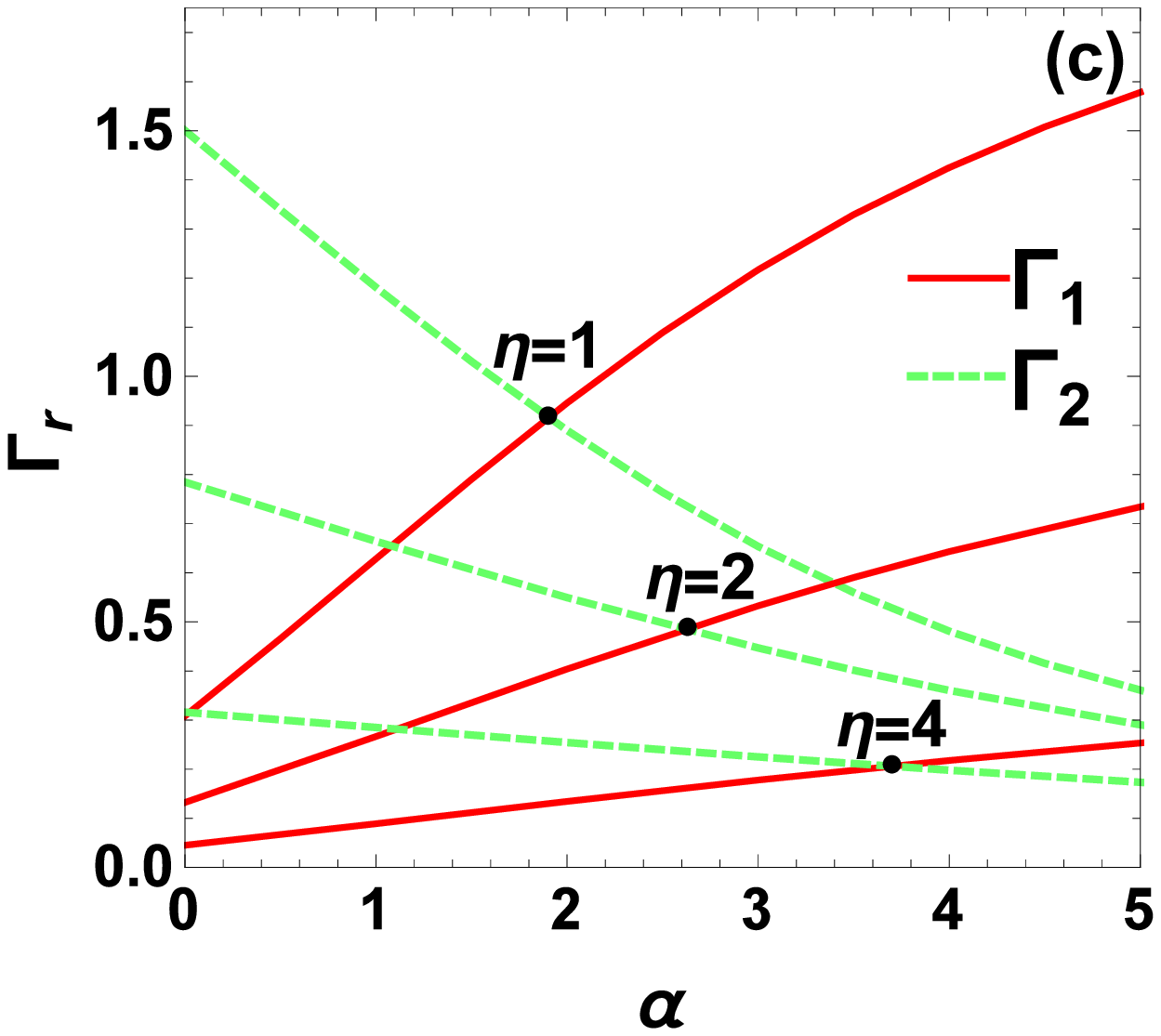}
\includegraphics[width=0.2389\textwidth]{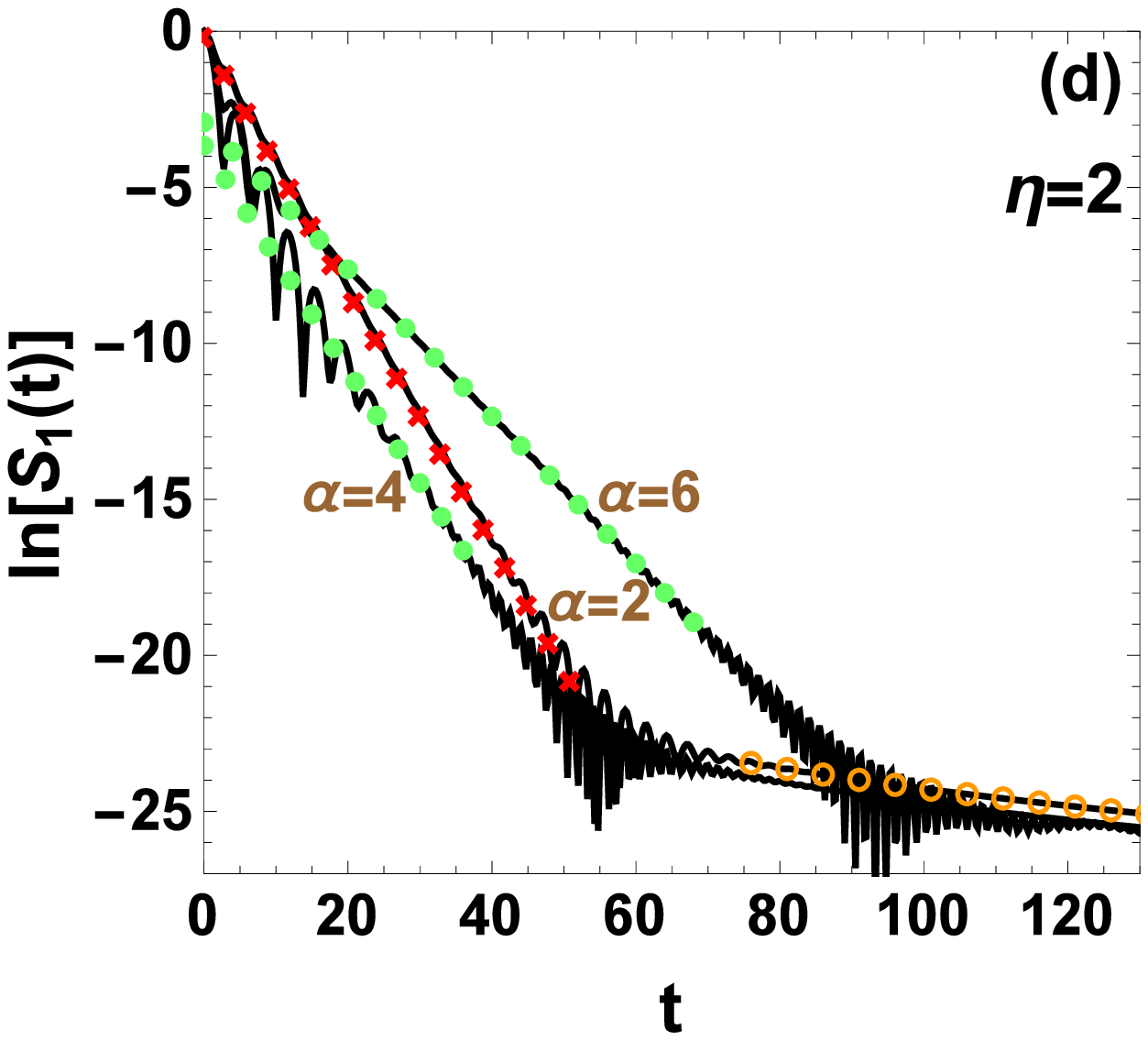}

\caption{Results for  the lowest-energy initial state, Eq. (\ref{initial1}).   (\textbf{\textit{a}}, \textbf{\textit{b}}, \textbf{\textit{d}})  Results  for $S_{1}(t)$ obtained for  different values of $\alpha$ and $\eta$ as functions of time $t$. The continuous  \textit{black} lines represent the  \textit{exact} numerical results.
  The   \textit{red} crosses  and  \textit{green} dots mark the results obtained from the one-pole approximations,  $S_{1}(t)\approx |M_{r }^{(1)}|^2e^{-\Gamma_{r} t}$, for $r=1$ and for $r=2$, respectively. The \textit{orange} circles represent  the  results obtained from the asymptotic formula of Eq.   (\ref{asym}).  (\textbf{\textit{c}}) Dependencies of  $\Gamma_{1}$  and $\Gamma_{2}$  as functions of $\alpha$ for   $\eta=1,2$ and   $\eta=4$.  The time is given in
$mL^2/\hbar$, strengths of the  barriers  in  $\hbar^2/ (m L) $  and  decay rates in  $\hbar/(mL^2)$.   \label{Fig1}}
\end{figure}

 \begin{figure}

\includegraphics[width=0.2389\textwidth]{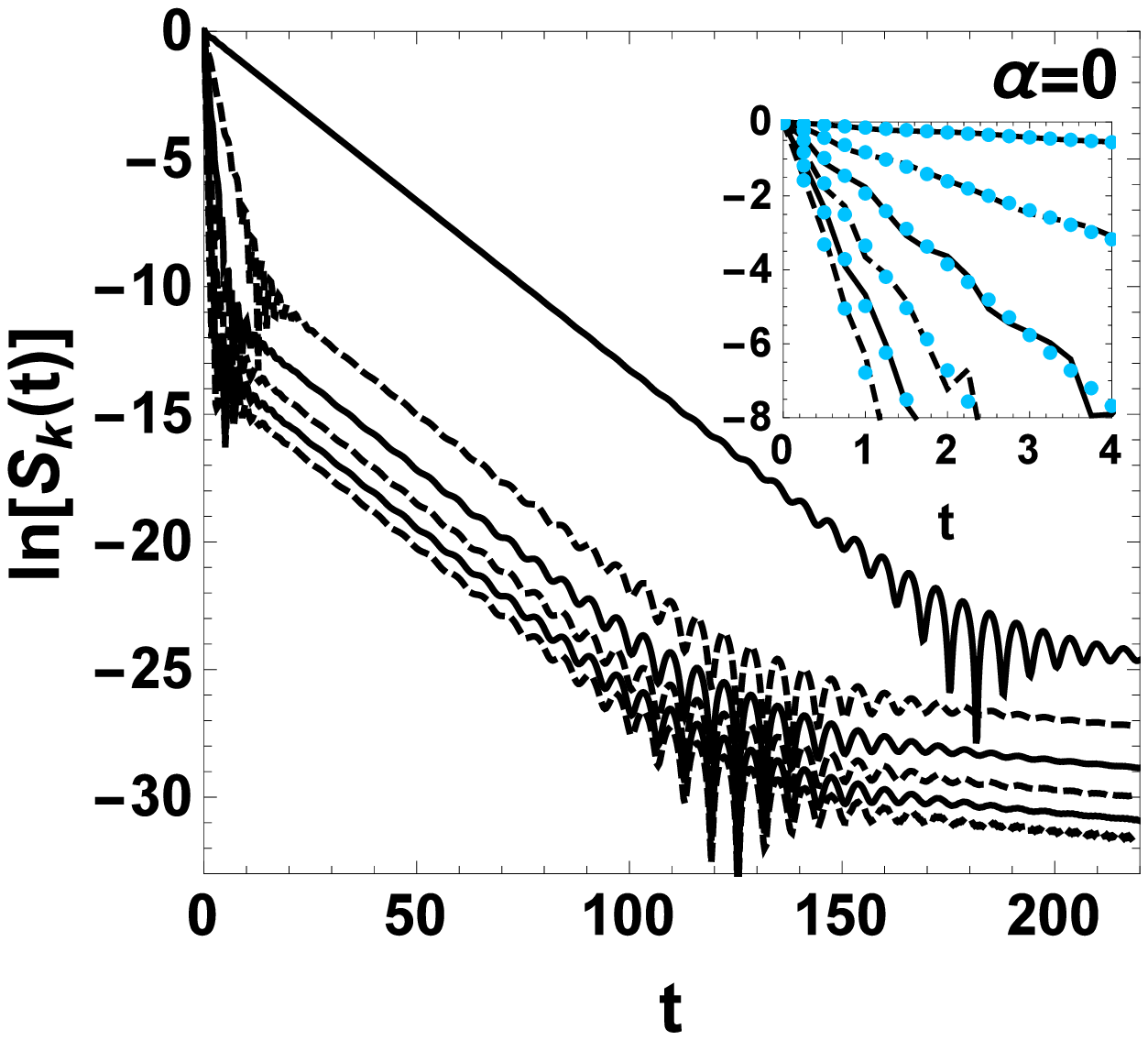}
\includegraphics[width=0.2389\textwidth]{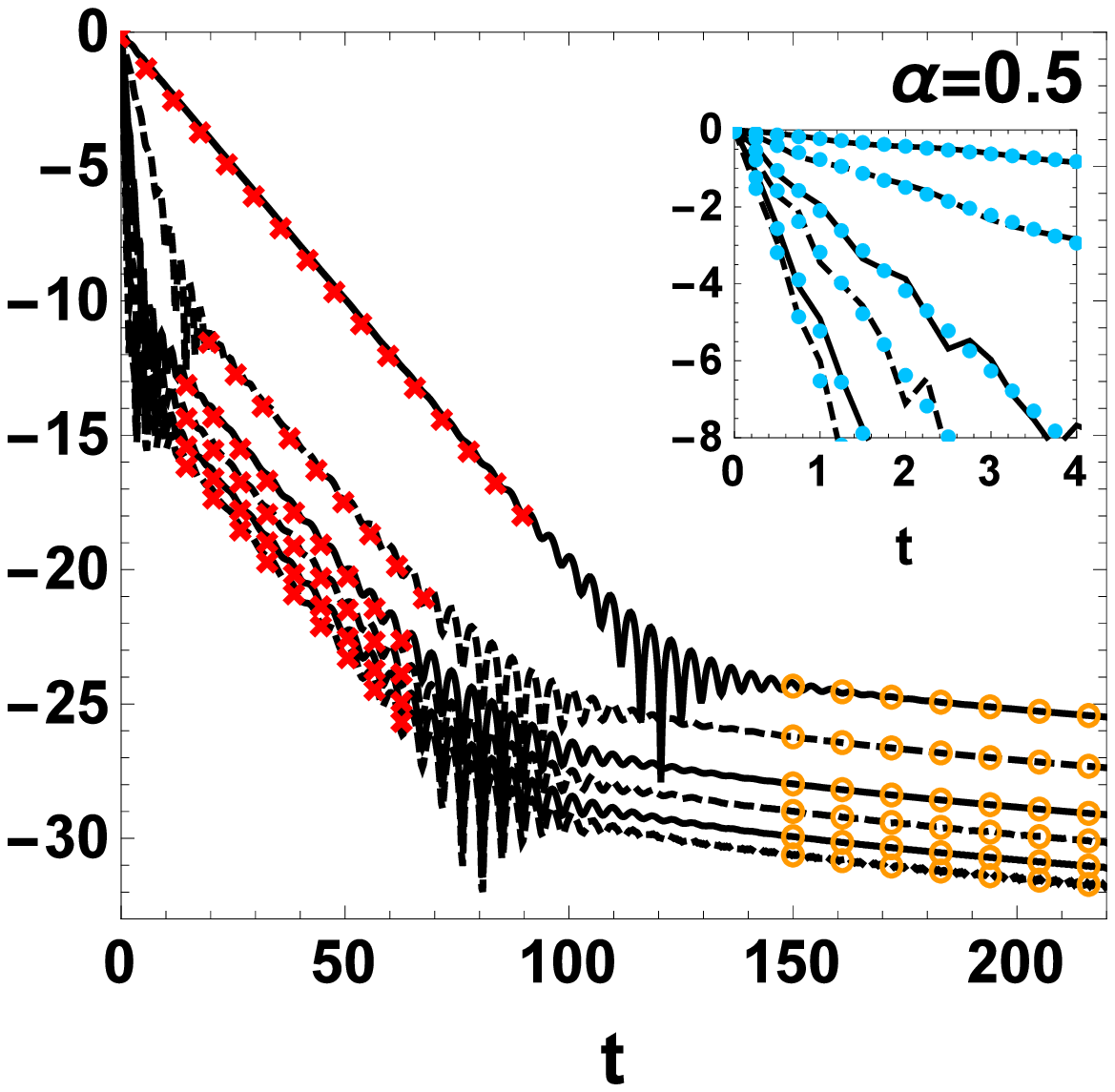}

\includegraphics[width=0.2389\textwidth]{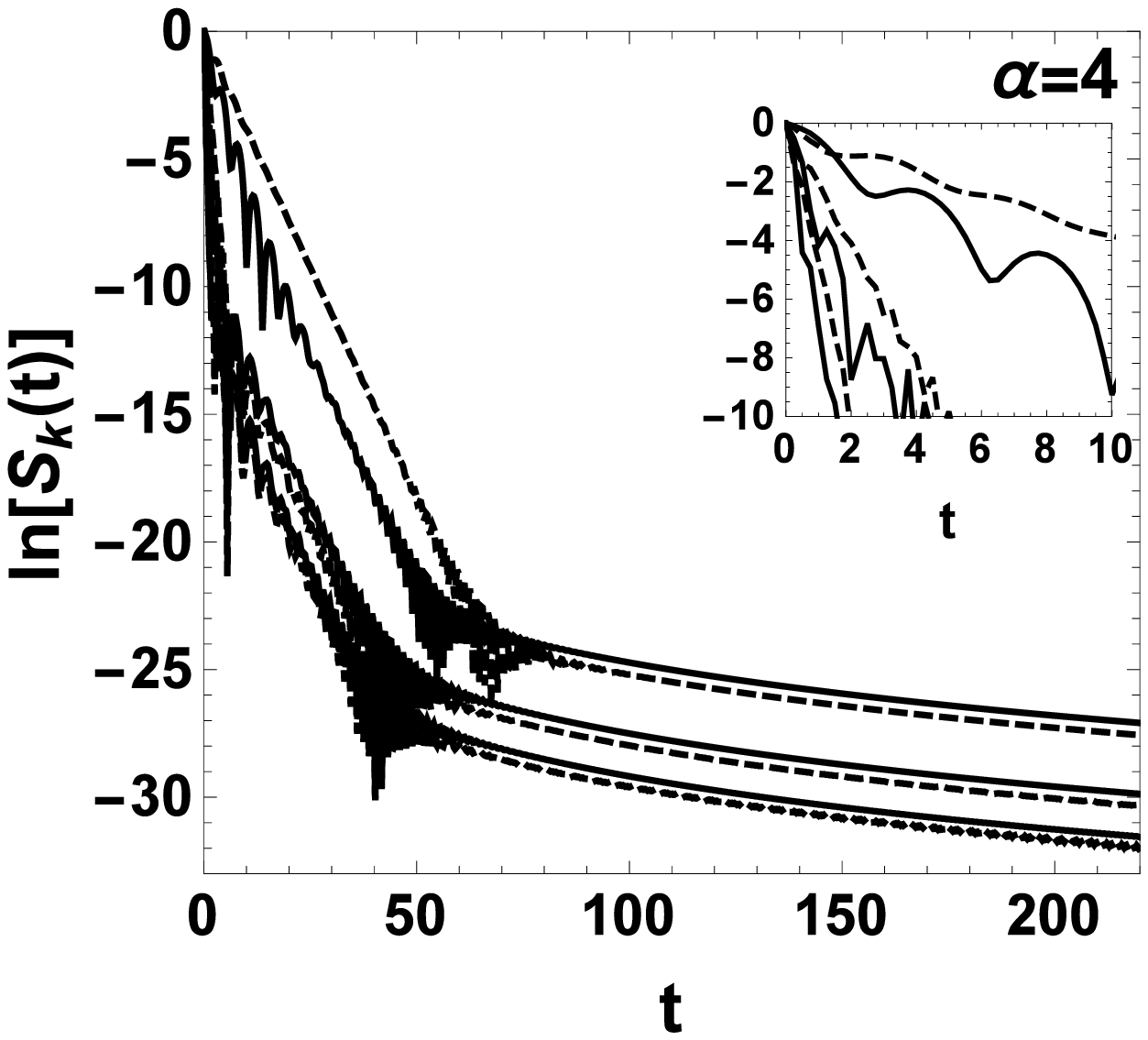}
\includegraphics[width=0.2389\textwidth]{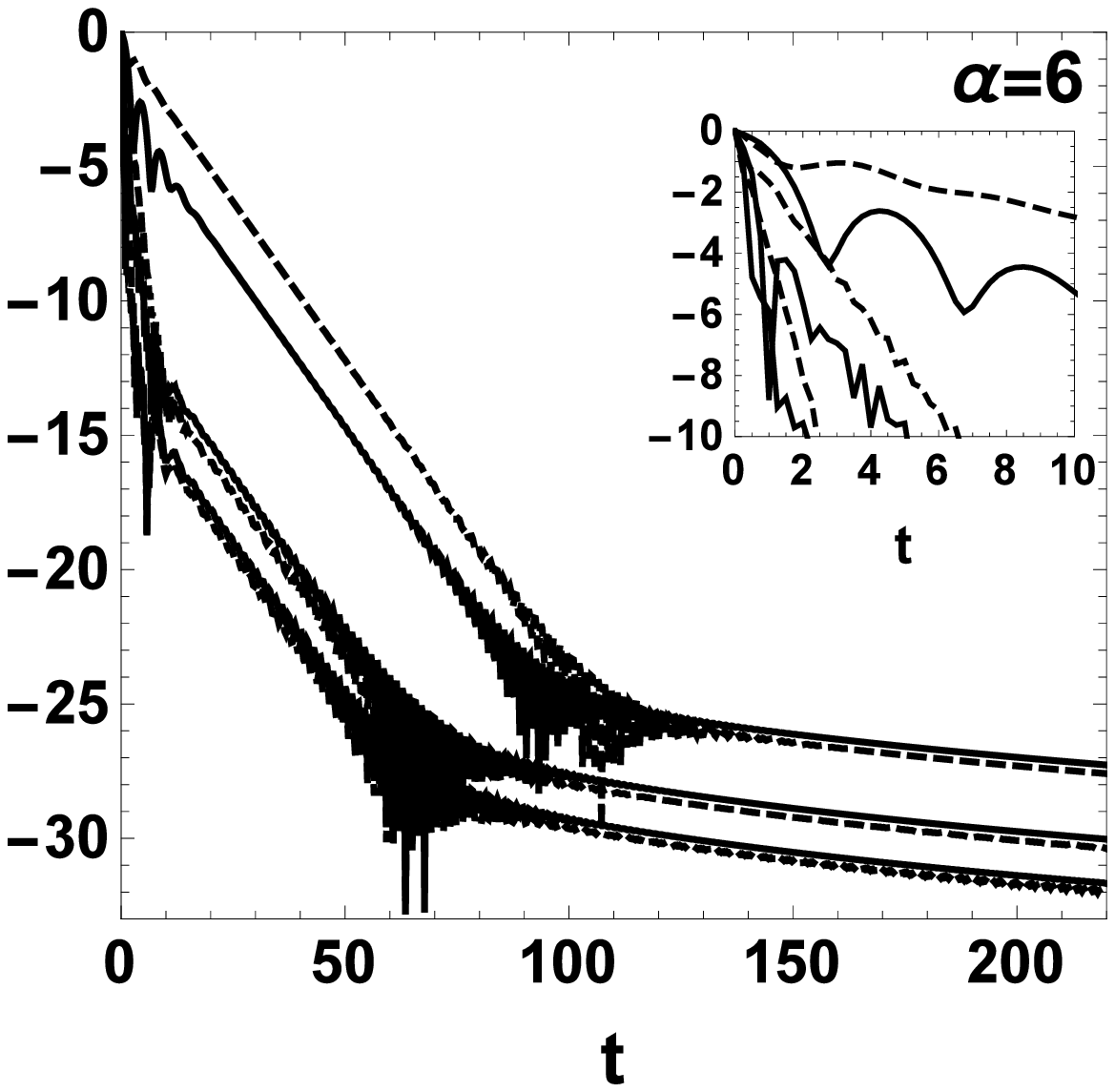}

\caption{Survival probabilities of six lowest-energy initial states obtained  for some transparent values of $\alpha$ at $\eta=2$. The continuous and dashed  \textit{black} lines  represent  the \textit{exact} results for  the even- and odd-parity initial states, respectively. In the plot of $\alpha=0.5$, the  \textit{red} crosses  and \textit{orange} circles mark the results obtained  from   $S_{k}(t)\approx |M^{(k)}_{1}|^2 e^{-\Gamma_{1}t }$ and  Eqs. (\ref{asym}) and (\ref{asym1}),
 respectively. The  \textit{blue}  dots  in the
insets show the results obtained from  $S_{k}(t)\approx e^{-\Gamma_{k} t}$ ($\alpha=0$ and $\alpha=0.5$). The  time and
 strength of the  barriers  are in $mL^2/\hbar$ and $\hbar^2/ (m L) $, respectively.    \label{Fig2}}
\end{figure}
First, we shed  light on the effect of a split barrier on the decay properties of one particle that is initially in the lowest energy state, Eq. (\ref{initial1}). Our results  are shown in Fig.  \ref{Fig1}, where   the \textit{exact} results for $S_{1}(t)$  are  obtained  by employing   the numerical estimation of the integrals    in Eq. (\ref{o}).
 An exponential decay begins in the early period. As  $\alpha$ increases, the slope of $\mathrm{ln}S_{1}(t)$ initially decreases, and when $\alpha$ exceeds some critical value, it starts to exhibit the opposite behavior.
To gain  deeper insights into this effect, we resort to the resonance expansion of the survival amplitude  in Eq. (\ref{amp1}).   Figure  \ref{Fig1}(\textbf{\textit{c}}) shows  the results obtained for the two lowest decay rates $\Gamma_{r}$ as functions of $\alpha$.  The curves intersect
 at a certain  point  $\alpha_{cr}$, which suggests that the following single-pole approximation holds:\begin{equation}\label{appo} S_{1}(t)\approx |M_{r }^{(1)}|^2e^{-\Gamma_{r} t},\end{equation}where $r=1$ for $\alpha<\alpha_{cr}$ and $r=2$ for $\alpha>\alpha_{cr}$.  Its applicability is verified in Figs. \ref{Fig1}(\textbf{\textit{a}}), \ref{Fig1}(\textbf{\textit{b}}),  and \ref{Fig1}(\textbf{\textit{d}}). For values of $\alpha$ differing considerably from    $\alpha_{cr}$, it  reasonably  reproduces the corresponding   \textit{exact}    results. For
$\alpha\approx\alpha_{cr}$, the resonance terms with $r=1,2$ in Eq. (\ref{amp1}) contribute substantially, as reflected by  the strong oscillations in the behavior of $S_{1}(t)$.
 The appearance of the transition point $\alpha_{cr}$  can be attributed to the fact that, at large enough values of $\alpha$, the average energy of the lowest decaying state approaches that of the first excited state. As a consequence, the decay energy spectrum   spreads more around the second resonance energy $\xi_{2}$, which corresponds to the  decaying fragment with $\Gamma_{2}$.
Strictly in the limit
  $\alpha \rightarrow\infty $, the    adjacent    even- and odd-parity initial  states [Eqs. (\ref{initial1}) and (\ref{initial2}), respectively] become degenerate, which means that their  decays proceed in the same way.

Figure \ref{Fig2} shows the behaviors of  the survival probabilities   for the six  lowest-energy initial  states.  It depicts how the results obtained for the even-parity states converge to those obtained for  the odd ones as    $\alpha$ is increased.
 In the      weak-split-barrier  regime,   all the states under consideration  undergo exponential decay in short times.
  The  profiles of $S_{k}(t)$  can be  reproduced well by $S_{k}(t)\approx e^{-\Gamma_{k} t}$ \cite{pons}, which is demonstrated   in the insets of  Fig. \ref{Fig2} (the cases $\alpha=0$ and  $\alpha=0.5$).
 As time progresses, the contributions from the poles gradually fade. As a result,  after a sufficiently long  period, the decays of the excited  states proceed with  the same decay constant as that of the lowest-energy  state \cite{TG0}. Then,   the following approximation applies:  $S_{k}(t)\approx |M^{(k)}_{r}|^2 e^{-\Gamma_{r}t}$,
  which   is confirmed with an example with    $\alpha=0.5$ ($r=1$) in Fig. \ref{Fig2}.  Eventually,  the decay of $S_{k}(t)$ follows a long-time inverse power law \cite{garcia}, which is derived in analytic form for both even- and odd-parity  initial states:

\begin{equation}\label{asym}
S^{(+)}_{k}\sim\frac{32\text{A}_{k}^4 \left \{\alpha^2  p_{k} \cos (p_{k})-
   \left [\alpha^2+(\alpha+1) p_{k}^2\right ] \sin (p_{k})\right \}^4}{\pi  \alpha^4 p_{k}^8
    [4 (\alpha+1) \eta+2 \alpha+1]^4t^3},
\end{equation}
and
\begin{equation}\label{asym1}S^{(-)}_{k}\sim\frac{32 (\alpha+1)^4}{\pi ^5 k^4 [4 (\alpha+1) \eta+2 \alpha+1 ]^4t^3},\end{equation}
 respectively. As may be seen  (Figs. \ref{Fig1}
and \ref{Fig2}), the results from  Eqs. (\ref{asym}) and (\ref{asym1}) accurately agree with the numerical ones, which confirms
their accuracy and correctness of our calculations.

     A feature worth stressing here is that in the regimes of values of $\alpha$  smaller (larger) than    $\alpha_{cr}$, the transition time to the long-time $t^{-3}$ regime decreases (increases) with an increase in $\alpha$. It can thus be concluded that by appropriately choosing the strength of the splitting barrier,  one can accelerate or slow down the decay process, keeping the strength of the trapping potential fixed at the same time.

\subsection{$N$-particle case}
\begin{figure}
\includegraphics[width=0.4\textwidth]{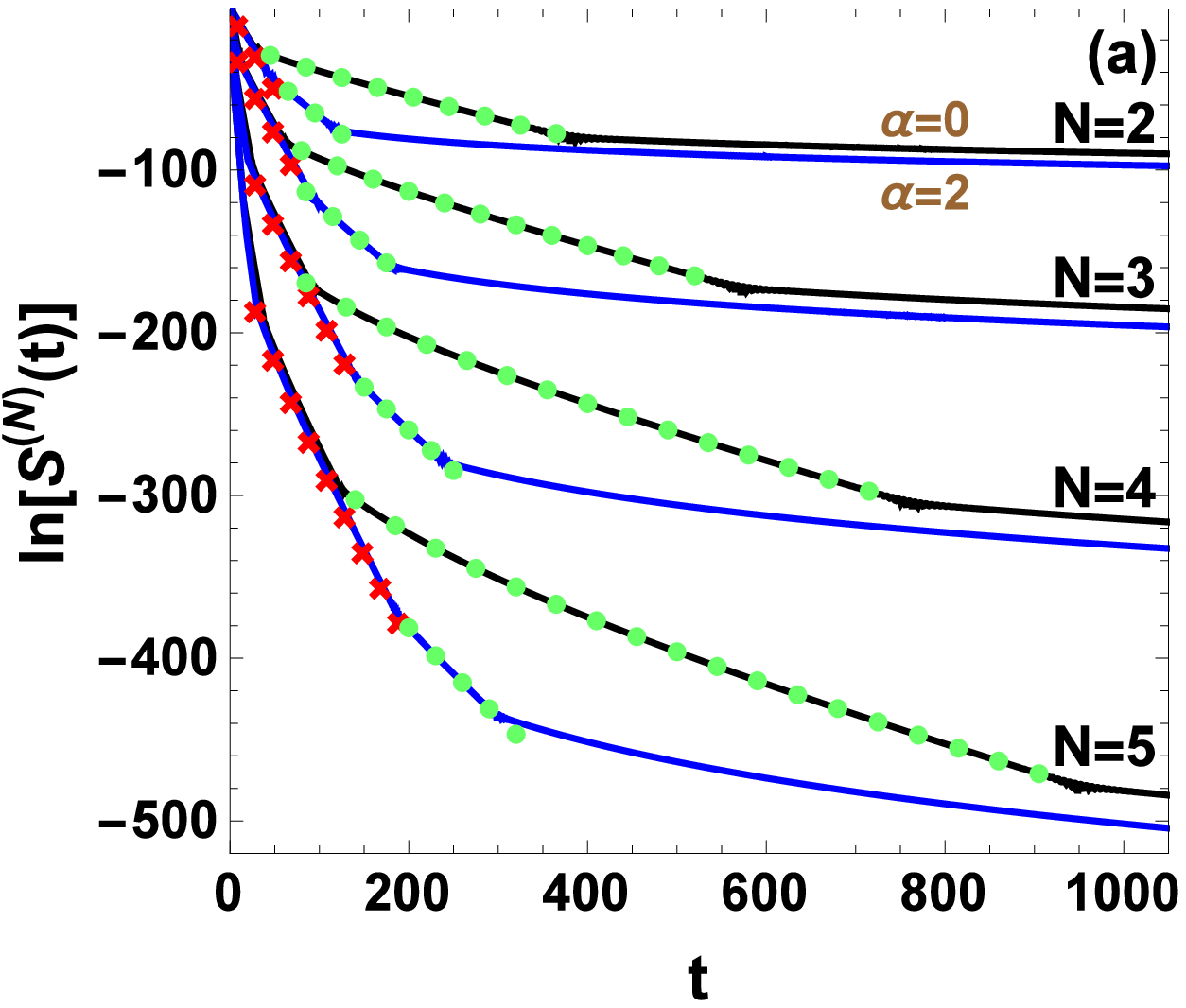}
\includegraphics[width=0.379\textwidth]{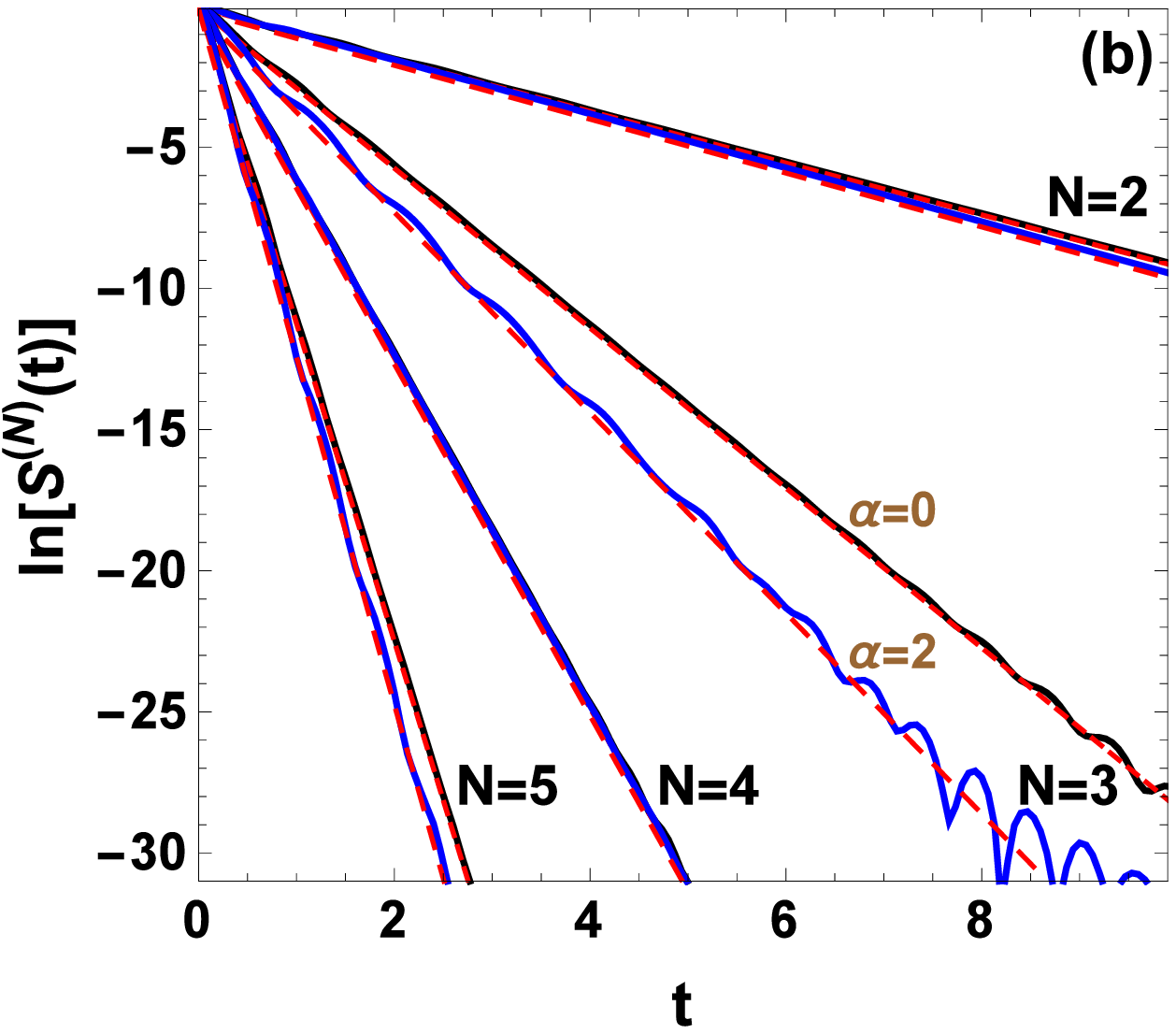}
\caption{Survival probabilities $S^{(N)}(t)$ obtained  at  $\eta=2$ for $\alpha=0$ (solid \textit{black}  lines) and  $\alpha=2$ (solid \textit{blue}  lines)  for up to   $N=5$.   (\textbf{\textit{a}}) The \textit{green} dots and \textit{red} crosses     mark the results obtained from Eqs. (\ref{reg3})  and  (\ref{reg2}), respectively.
  (\textbf{\textit{b}}) The dashed \textit{red} lines show the   results obtained  from Eq. (\ref{ap9}).  The time and
 strength of the  barriers  are in  $mL^2/\hbar$ and $\hbar^2/ (m L) $, respectively.   \label{Fig3}}
\end{figure}Now, we come to the main point, where we explore the decay properties of the  TG gases  \cite{Girardeau}.  Below,  we restrict our investigation to cases with sufficiently low split barriers without losing much generality.
 Our  results  for the $N$-particle survival probability  $S^{(N)}(t)$ are   summarized in    Fig. \ref{Fig3}, where, for the sake of clarity,  we only  depict   the results for      $\alpha=0$ and $\alpha=2$.    Figure \ref{Fig3}(\textbf{\textit{a}}) shows  the results   obtained   using   the resonance expansion approach. In contrast,   the results of  Fig. \ref{Fig3}(\textbf{\textit{b}}) highlight
 the changes of $S^{(N)}(t)$ in a short-time regime, and they   were obtained by direct numerical integration in Eq. (\ref{o}).

We can identify several distinct regions with   different behaviors
of $S^{(N)}(t)$ that  deserve a  profound investigation.
To this end, we use a single-term approximation extracted from the resonance expansion of $S^{(N)}(t)$ in the following form:
\begin{equation}\label{reg0}S^{(N)}(t) \approx {\cal N}_{n} e^{-\Gamma^{(n)}t}t^{-\beta},\end{equation}
with  \begin{equation}\Gamma^{(n)} =\sum_{k=1}^{n} \Gamma_{k}, n\leq N,\end{equation}
and the smallest exponent $\beta$, which we denote by $\beta(n,N)$.
  A close inspection reveals   that  $\beta(n,N)$ is in a simple relation with  an exponent in  the long-time asymptotic law for the TG system
  \cite{TG1}:  \begin{equation}\label{long}S^{(N)}(t)\propto t^{-\gamma(N)},\end{equation}$\gamma(N)= N(2N+1)$, namely, $\beta(n,N)= \gamma(N-n)$.
   In this context,
    see also the informative  discussion in Ref. \cite{TG1}.   Equation (\ref{reg0}) can be expected     to  hold in  cases when the true $S^{(N)}(t)$ resembles an exponential decay.  For the reasons  mentioned below Eq. (\ref{o}), the   formula  in   Eq. (\ref{reg0}) also  holds  for the $N$-particle nonescape probability (only the factors ${\cal {N}}_{n}$ are different). However, it is important to stress that,   in contrast to  Eq. (\ref{long}),
 Eq.  (\ref{reg0}) must be regarded as approximate.

Let us  begin with the approximation for  $n=N$:   \begin{equation}\label{ap9}S^{(N)}(t)\approx  {\cal N}_{N}e^{-\Gamma^{(N)} t}.\end{equation}
     Equation (\ref{ap9}) has already appeared before in Ref. \cite{pons} and is valid to some extent, as confirmed by the results of   Fig. \ref{Fig3}(\textbf{\textit{b}}). Although, at a short period, the $S^{(N)}(t)$   are only slightly dependent on $\alpha$,      the effect of particle number is noticeable. To understand this phenomenon,  we refer again to Fig.  \ref{Fig1}(\textbf{\textit{c}}). Its inspection reveals that the sum of the two lowest decay rates $\Gamma_{r}$, that is, $\Gamma^{(2)}$,  does not change much as $\alpha$ is varied. The presence of the additional particle
  prevents this effect, which manifests itself in the appearance of a  stronger dependence of $\Gamma^{(3)}$  on $\alpha$.
The same  mechanism is responsible  for the behaviors of  $S^{(N)}(t)$
  for  a larger particle number $N$ as the results of Fig. \ref{Fig3}(\textbf{\textit{b}}) clearly indicate.  It follows from the above that Eq. (\ref{reg0}) is much more sensitive to changes in $\alpha$ for odd values of $n$ than for the even ones.

Next, we restrict ourselves to  two particular time regions (long-lived) appearing  before the long-time asymptotic regime. It can be expected that, in these regions,  the $S^{(N)}(t)$   behave in accordance  with  Eq. (\ref{reg0}) as follows:
 \begin{equation}\label{reg3}S^{(N)}(t) \approx {\cal N}_{1} e^{-\Gamma^{(1)}t} t^{- ( 2 N-3) N-1}, \end{equation}
and
\begin{equation}\label{reg2}S^{(N)}(t) \approx {\cal N}_{2} e^{-\Gamma^{(2)}t} t^{ -( 2 N-7) N-6}. \end{equation} The first equation corresponds to the regime before  to the transition to the  long-time $t^{-\gamma(N)}$ regime. Note  that for  $N=2$, Eq. (\ref{reg2})  reduces to Eq. (\ref{ap9}). The results obtained from the above  approximations  are plotted along with the \textit{exact} results in Fig. \ref{Fig3}(\textbf{\textit{a}}). For the clarity of the presentation,  we display  the results   of Eq.  (\ref{reg2})  for only the case of $\alpha=2$ (we recall that $\Gamma^{(2)}$ is practically insensitive to changes in $\alpha$).    Our results show  an excellent agreement between the predictions of the above approximations and the     \textit{exact} results, thereby confirming the validity of Eq.  (\ref{reg0}) and  disclosing the nature of the decay process of TG gases.
  It results from our study   that only the initial phase  of  the decay of TG gases  may follow   an exponential law. The effect of the split barrier is most visible   before the long-time asymptotic regime, which corresponds to Eq.  (\ref{reg3}).  As far as we know,  in none of the publications has Eq. (\ref{reg0}) appeared.

Finally, we  analyze the fraction  of particles within the trap, which is now  easily  accessible by  experimental observation techniques \cite{johim2}:
\begin{equation}\label{ava}{N}_{T}(t)=\int_{-1}^{1} \rho(x,t)\mathrm{d}x,\end{equation} where $\rho(x,t)$ is the one-body density matrix normalized to the number of particles $N$. Similarly,
the fraction of particles in  the left and  right wells can be defined as    ${N}_{L}(t)=\int_{-1}^{0} \rho(x,t)\mathrm{d}x$ and ${N}_{R}(t)=\int_{0}^{1} \rho(x,t)\mathrm{d}x$, respectively.
\begin{figure}
\includegraphics[width=0.242\textwidth]{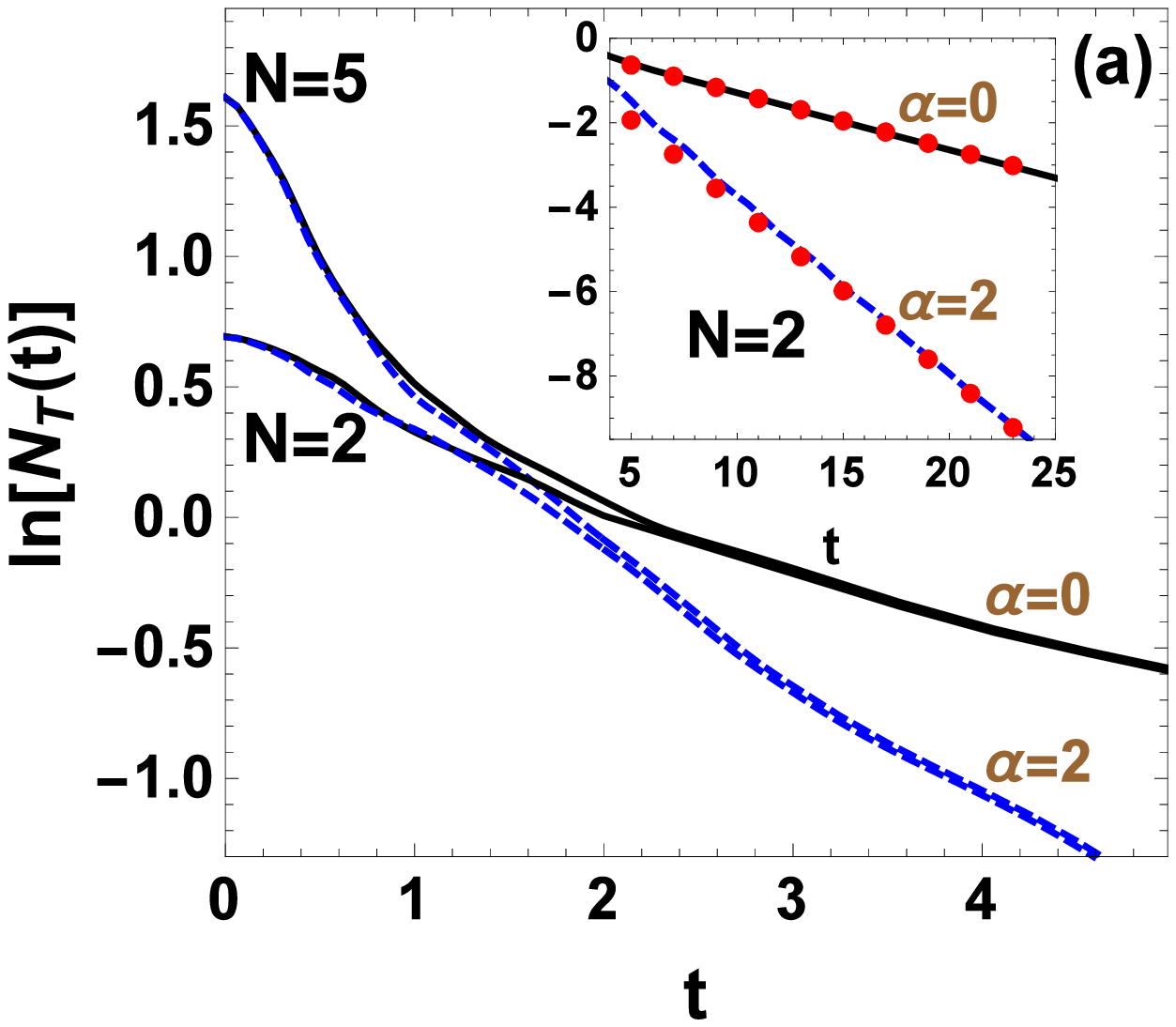}
\includegraphics[width=0.235\textwidth]{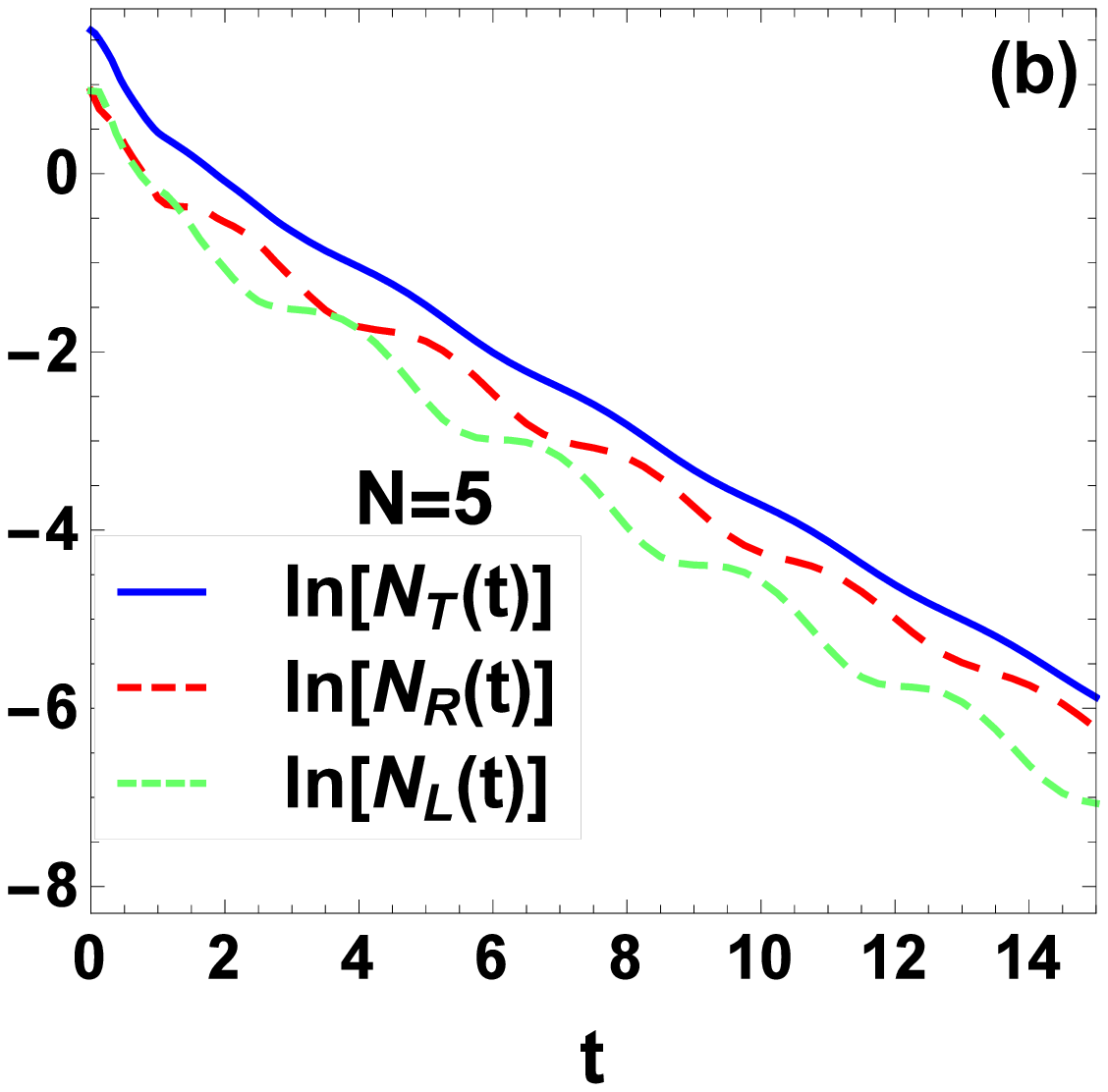}
\caption{  (\textbf{\textit{a}}) Results for the fraction of particles  in the trap  ${N}_{T}(t)$
  obtained for the systems of $N=2$ and $N=5$  at  $\eta=2$ for $\alpha=0$ (solid \textit{black}  lines) and  $\alpha=2$ (dashed \textit{blue}  lines).    The inset compares  the results obtained  from the resonance expansion approach (\textit{red} dots, Eq. (\ref{eq22})) with the numerical ones.  (\textbf{\textit{b}}) Corresponding results obtained  for  ${N}_{L/R}(t)$ at $N=5$ and $\alpha=2$. The time and
 strength of the  barriers  are in  $mL^2/\hbar$ and $\hbar^2/ (m L) $, respectively.   \label{Fig4}}
\end{figure}

 For the TG systems, Eq. (\ref{ava})  simplifies to the form
${N}_{T}(t)=\sum_{i=1}^{N}\rho_{i}(t)$, where       $\rho_{i}(t)=\int_{-1}^{1} |\phi_{i}(x,t)|^2\mathrm{d}x$. The time evolution of  ${N}_{T}(t)$      was  studied    as a function of   $N$ in Ref. \cite{TG0} in the absence of  a split barrier.     Therefore,  we only focus   on  the  effect of the split barrier,  limiting ourselves to systems with $N=2$ and $N=5$.
Figure \ref{Fig4} (\textbf{\textit{a}}) shows our numerical results for ${N}_{T}(t)$      calculated for the same control parameter values of the trap  as those in Fig. \ref{Fig3}.  Figure \ref{Fig4}(\textbf{\textit{b}}) displays the corresponding results for  ${N}_{L/R}(t)$ obtained for  $N=5$ at $\alpha=2$.
The time evolution of ${N}_{T}(t)$    can easily be   understood from the facts established     for  the decay   of  one-particle states.      At the beginning of the time period, the behaviors of $\rho_{i}(t)$ can be identified as   consistent with the approximation:  $\rho_{i}(t)\approx e^{-\Gamma_{i}t}$. Then,  the quantity ${N}_{T}(t)$ is almost insensitive to changes in   $\alpha$,  as shown in Fig.  \ref{Fig4}(\textbf{\textit{a}}).
By contrast, ${N}_{T}(t)$ experiences an exponential decay   when its components  $\rho_{i}(t)$ go into the behaviors  $\rho_{i}(t)\approx n_{i}e^{-\Gamma_{1}t}$, i.e.,
    \begin{equation}\label{eq22}  {N}_{T}(t)\approx {\cal N}_{t}e^{-\Gamma _{1}t},\end{equation} where  ${\cal N}_{t}=\sum_{i=1}^{N} n_{i}$. The validity of this approximation
 is verified both  with and without the  split barrier  in    the inset of Fig.\ref{Fig4}(\textbf{\textit{a}}), where  the  values of  $n_{i}$ were determined with   the   resonance  expansion approach. We conclude that the presence of the split barrier primarily has a major impact on the exponential decay of ${N}_{T}(t)$. However, contrary to the behaviors of  ${N}_{T}(t)$, those  of  $ {N}_{L/R}(t)$ cannot be regarded  as  exponential.    The slopes of   $\mathrm{ln} {N}_{L/R}(t)$  display evident oscillations,
which reflects the fact that    resonance  terms other than $  e^{- \Gamma_{1}t}$  contribute  considerably  to ${N}_{L/R}(t)$.
  This phenomenon    can be  reasonably  attributed   to the effects of  the particle tunneling between the   wells.

\section{Conclusion}\label{section3}
In conclusion, we have carried out a detailed study of
the decay properties of   TG gases escaping from a double well modeled by  \textit{Winter's system} with the center $\delta$-split barrier. We provided an explicit representation of the time-dependent single-particle states in terms of the Fourier integrals, enabling accurate evolutions of the  TG wave functions with time. Using  the resonance expansion approach, we  derived the closed-form approximate expression for the survival probability in the general case of $N$ particles. We demonstrated its validity at different stages of the time evolution, thus revealing the mechanism of  the decay process of unstable TG gases. We  also explained the parity effect appearing in the behavior of the survival probability in the presence of a split barrier.  Our study  establishes that only in the initial   period can the decay of  TG gases be regarded as purely exponential.
In addition, we investigated the effect of the split barrier on    the fraction of particles in the trap during the time evolution.

We leave open the question of how the features revealed in this study are affected if one considers the finite interaction strength and/or models with other confining potentials.The decay of many-particle systems  exhibits an exponential behavior   for any   interaction strength (e.q., Ref. \cite{Lode}); however,  little is known about    its  course  at further stages of the time evolution.
 In the finite interaction case,   the classical  counterpart of the system can have a dynamical instability. As a result,  one can expect significant deviations from the decay features established in this paper for     TG gases  \cite{eho,eho1}.
Another route for further study   is to investigate in detail  the effect    of  particle tunneling      between   the left and right wells.

 We believe that our study will  broaden discussion about the many-particle decay process.

\section{Appendix}
\subsection{Continuous spectrum eigenfunctions}\label{ap1}
To find the time dependence of $\phi_{k}(x,t)$  we solved the Schr\"{o}dinger equation (\ref{sh}) for the eigenstates of
  the continuous spectrum.
    The   problem is equivalent to
  a free particle  provided  that  its  wave function is continuous in the whole space and  satisfies specific  matching  conditions at the positions  of the  $\delta$ barriers \cite{cont}:
$$\lim_{\epsilon\rightarrow 0^{+}} [\psi^{'}(\epsilon)-\psi^{'}(-\epsilon)]=2\alpha\psi(0),$$
$$\lim_{\epsilon\rightarrow 0^{+}} [\psi^{'}(1+\epsilon)-\psi^{'}(1-\epsilon)]=2\eta\psi(1).$$
 We omit here  the derivation details and report our final step:

\begin{equation}\label{fun}
{\cal \psi}_{p}(x) =\mathrm{A}(p) \left\{
\begin{array}{lll}
 \sin (p x)+\tan (p) \cos(px),  -1\leq x \leq 0 \\
\mathrm{B}_{p} \sin (p x)+\tan (p) \cos (px) , 0<x<1, \\\mathrm{C}_{p} \sin (p x)+\mathrm{D}_{p} \tan (p)\cos (p x),x \geq 1
\end{array} \right.
\end{equation}
where

\begin{equation}
\mathrm{B}_{p}=\frac{2 \alpha  \tan (p)}{p}+1,
\end{equation}

\begin{equation}
\mathrm{C}_{p}=2\frac{\alpha\eta[1-  \cos (2 p)]+ \alpha p \tan (p)+ \eta p \sin (2 p)}{p^2}+1,
\end{equation}

\begin{equation}
\mathrm{D}_{p}= 2\frac{ \alpha \eta [\cos (2 p)-1]- \eta p \sin (2 p)}{p^2}+1,
\end{equation}
with $\mathrm{A}(p)$ determined by  a standard normalization for the continuous spectrum,
\begin{equation}
\int_{-1}^{\infty}  {\cal \psi}_{p^{'}}(x){\cal \psi}_{p}(x) \mathrm{d}x=\delta(p^{'}-p),
\end{equation} that is,
\begin{widetext}
\begin{equation}
 \mathrm{A}(p)={\sqrt{\frac{2}{\pi }} p^2\over \sqrt{4 \tan (p) [\alpha \tan (p)+p] \left \{\alpha \left(2 \eta^2+p^2\right)+2 \eta \left [\cos (2 p) \left(p^2-\alpha \eta\right )+p (\alpha+\eta) \sin (2 p)\right ]\right\}+p^4/ \cos ^2(p)} }.
\end{equation}\end{widetext}
where $p=(2E)^{1/2}$.

\subsection{Eigenfunctions of a hard- wall split trap }\label{ap2}
Even-parity $(+)$  one-particle eigenfunctions  of the hard-wall split trap
  are found as \cite{exact}
  \begin{eqnarray}\label{initial1}
{ \phi_{k}}^{(+)}(x,0) =\mathrm{A}_{k} \left\{
\begin{array}{ll}
  \cos(p_kx)- \frac{\alpha}{p_k}\sin (p_{k} x),  -1\leq x \leq 0 \\
 \cos (p_{k}x)+\frac{\alpha}{p_k}\sin (p_{k} x) , 0<x\leq 1
\end{array} \right.,
\end{eqnarray}
where $\mathrm{A}_{k}$ is the normalization factor and
the values of $p_{k}$  are  the solutions of the transcendental equation:   $\tan(p)=-p/\alpha$. The energies  are given by $E_{k}^{(+)}=p_{k}^2/2$ and ordered so as to form an increasing sequence.
In the odd-parity case  $(-)$ the wave functions vanish at $x=0$ and  therefore   are  the same as those for the pure hard-wall trap; that is,
\begin{equation}\label{initial2}{ \phi_{k}}^{(-)}(x,0)=\sin (\pi  k x), \end{equation}
with  energies   $E_{k}^{(-)}=(\pi  k)^2/2$.

\subsection{Expansion coefficients $C_{k}(p)$}\label{ap3}

  In both Eqs. (\ref{initial1}) and (\ref{initial2}), the integral in Eq. (\ref{mom}) can be performed algebraically:

\begin{widetext}\begin{equation}\label{mome} C_k^{(+)}(p)=\frac{2 \mathrm{A}_{k} \mathrm{A}(p) \sin (p) \left \{p \left(\alpha^2+{p}_{k}^2\right) \sin ({p}_{k})-\tan
   (p) \left [{p}_{k} \left(\alpha^2+p^2\right) \cos ({p}_{k})+\alpha (p^2-{p}_{k}^2)  \sin
   ({p}_{k})\right]\right \}}{\alpha p (p^2-{p}_{k}^2) },\end{equation}\end{widetext}
and
\begin{equation}\label{mome1}C_{k}^{(-)}(p)=\frac{2\mathrm{A}(p) \pi   k (-1)^k  \sin (p) (\alpha \tan (p)+p)}{p (p^2-\pi^2  k^2) },\end{equation}
 respectively.  The coefficients $C_{k}(p)$ and the related quantities are ordered according to the energy spectrum of  the  split  hard-wall trap as
$C_{k}(p)=C^{(-)}_{{k/ 2}}(p) $ and $C_{k}(p)=C^{(+)}_{{k/2+1/2}}(p)$ for  even and odd values of  $k$, respectively.

\section{ACKNOWLEDGMENTS}
The author is thankful to Tomasz Sowi\'{n}ski and Arkadiusz Kuro\'{s}  for their comments and suggestions on a draft of this article.

\end{document}